# Feynman's Interpretation of Quantum Theory[1]

H. D. Zeh

www.zeh-hd.de

**Abstract:** A historically important but little known debate regarding the necessity and meaning of macroscopic superpositions, in particular those containing different gravitational fields, is discussed from a modern perspective.

## 1. Introduction

Richard Feynman is most famous for his unprecedented mastery in applying the quantum theory to complex situations, such as quantum field theory and quantum statistical mechanics, by means of novel and mainly intuitive methods and concepts. However, he is also known for his remark "I think I can say that nobody understands Quantum Mechanics." So he evidently distinguished between being able to use a theory and understanding it.

Let me, therefore, first point out that Feynman explained on several occasions that he had originally hoped his path integral formalism[1] would represent a new and possibly self-explanatory quantum theory, but that he soon had to realize (not least because of Dyson's arguments[2]) that this formalism was but a new method to calculate the propagation of wave function(al)s for particles and fields in the Schrödinger picture. This will also become evident in the discussion that is quoted below. It thus appears inappropriate to use the path integral formalism in an attempt to justify unitarity.[3] In particular, Feynman's famous graphs, which seem to contain particle lines, are exclusively used as an intuitive means to construct terms of a perturbation series, where the particle lines are immediately replaced by plane waves or free field modes appearing under an integral. His paths in configuration space, on the other hand, are often entirely misunderstood as forming *ensembles* (claimed to be required by Heisenberg's uncertainty relations), from which subensembles could then be picked out by merely increasing one's knowledge. Such a "conventional" statistical interpretation of quantum mechanical superpositions is sharply rejected by Feynman in the following discussion, when he answered Bondi's comparison of quantum measurements with throwing dice.

As far as I know, Feynman never participated in the published debate about interpretational problems, such as quantum measurements. So I was surprised when I recently disco-vered a little known report about a conference regarding the rôle of gravity and the need for its quantization, held at the University of North Carolina in Chapel Hill in 1957,[4] since it led at some point to a discussion of the measurement problem and of the question about the existence and meaning of macroscopic superpositions. [Note added after completion: The report is now also available elsewhere[5] in a slightly revised form – see http://www.edition-open-access.de/downloads/files/Sources_5_published_V1.pdf. My quotations, which were from pp. 135-140, 149, 150, and 154 of Ref. 4, can there be found on pp. 249-256, 270, 272, and 278, respectively.] This session was dominated by Feynman's presentation of a version of Schrödinger's cat, in which the cat with its states of being dead or alive is replaced by a macroscopic massive ball being centered at two different positions with their distinguishable gravitational fields.

---

[1] Dedicated to the late John A. Wheeler – mentor of Richard Feynman, Hugh Everett, and many other great physicists.

I found this part of the report so remarkable for historical reasons that I am here quoting it in detail for the purpose of discussing and commenting it from a modern point of view. Let me emphasize, though, that one has to be careful when drawing conclusions about Feynman's or other participants' true and general opinions on the matter, since the report, edited by Cecile DeWitt-Morette, is according to her Foreword partly based on tape recordings, and partly on notes taken by herself and others, or provided by the participants. Some remarks may furthermore crucially depend on the specific circumstances of the conference. However, the text appears very carefully prepared and consistent, so I will here take it for granted. The discussion to be quoted below certainly deserves to become better known and discussed because of the influence it seems to have had on several later developments. Because of its spontaneous nature, I have decided to present most of my comments in a similar form of spontaneous remarks – even though they are fifty years late!

**2. Commented excerpts from Session 8 of WADC TR 57-216**

I shall begin on page 135 of Cecile DeWitt's report with two contributions which directly preceded Feynman's first remark in this context. This session of the conference had started with several other contributions on various subjects, in particular the meaning and validity of the equivalence principle in quantum gravity. Quotations from the report are indented in order to make them easily readable independently of my comments:

> **Salecker** then raised again the question why the gravitational field needs to be quantized at all. In his opinion, charged quantized particles already serve as sources for a Coulomb field which is not quantized. [Editor's note: Salecker did not make completely clear what he meant by this. If he meant that some sources could be repre-sented by actions-at-a-distance, then, although he was misunderstood, he was right. For the corresponding field can then be eliminated from the theory and hence remain unquantized. He may have meant that to imply that one should try to build up a completely action-at-a-distance theory of gravitation, modified by the relativistic necessities of using both advanced and retarded interactions and imbedded in an "absorber theory of radiation" to preserve causality. In this case, gravitation *per se* could remain unquantized. However, these questions are not discussed until later in the session.]

This question of whether quantization must be applied to the electromagnetic field or only to its sources dates back to Max Planck's proposal of the quantum of action. It is still under dispute today, though mainly for the (topological) *Coulomb constraint* (Gauss's law), while the vector potential represents dynamical degrees of freedom that must be quantized (see the further discussion). The kinematical Coulomb constraint, too, could alternatively be understood as a *dynamical* (retarded *or* advanced) causal consequence of charge conservation.

**Belinfante** insisted that the Coulomb field *is* quantized through the $\psi$-field. He then repeated DeWitt's argument that it is not logical to allow an "expectation value" to serve as the source of the gravitational field. There are two quantities which are involved in the description of any quantized physical system. One of them gives information about the general dynamical behavior of the system, and is represented by a certain operator (or operators). The other gives information about our knowledge of the system; it is the state vector. Only by combining the two can one make predictions. One should remember, however, that the state vector can undergo a sudden change if one makes an experiment on the system. The laws of Nature therefore unfold continuously only as long as the observer does not bring extra knowledge of his own into the picture. This dual aspect applies to the stress tensor as well as to everything else. The stress tensor is an operator which satisfies certain differential equations, and therefore changes continuously. It has, however, an expectation value which can execute wild jumps depending on our knowledge of the number and behavior of mass particles in a certain vicinity. If this expectation value were used as the source of the gravitational field then the gravitational field itself – at least the static part of it – would execute similar wild jumps. One can avoid this subjective behavior on the part of the gravitational field only by letting it too become a continuously changing operator, that is, by quantizing it. These conclusions apply at least to the static part of the gravitational field, and it is hard to see how the situation can be different for the transverse part of the field, which describes the gravitational radiation.

The "static part" of a classical field would not be well defined if the source were accelerating. The "$\psi$-field" here seems to be meant to describe matter only (such as individual electrons). Quantum states of matter and radiation would in general be entangled – a consequence of Schrödinger's theory that was not yet sufficiently appreciated at the time of the conference, even though entanglement between distant systems had been discussed ever since the paper of Einstein, Podolski and Rosen of 1935 had become known.

Belinfante's description appears typical for the Heisenberg picture, and in this way, as we shall see, forms an illustrative contrast to Feynman's understanding of quantization and the rôle of the wave function. The click of a counter, for example, can hardly be attributed to a sudden increase of our knowledge – although it may *cause* such an increase. Belinfante, who is known for supporting hidden variables,[6] here clearly understands the wave function as an epistemic concept (corresponding to Heisenberg's "human knowledge"), so it must change for reasons beyond the system's physical dynamics. He does *not* refer to ensembles of wave functions or a density matrix in order to represent incomplete knowledge. Note, however, that Belinfante is only using the word "knowledge" in an epistemic sense, while his "information" seems to refer to an objective representation or description (of dynamics or knowledge).

**Feynman** then made a series of comments of which the following is a somewhat condensed but approximately verbatim transcript:

"I'd like to repeat just exactly what Belinfante said with an example – because it seems clear to me that we're in trouble if we believe in quantum mechanics but *don't* quantize gravitational theory. Suppose we have an object with spin which goes through a Stern-Gerlach experiment. Say it has spin 1/2, so it comes close to one of two counters.

This text is accompanied by a simple schematic drawing.

Connect the counters by means of rods, etc., to an indicator which is either up when the object arrives at counter 1, or down when the object arrives at counter 2. Suppose the indicator is a little ball, 1cm in diameter.

"Now, how do we analyze this experiment according to quantum mechanics? We have an amplitude that the ball is up, and an amplitude that the ball is down. That is, we have an amplitude (from the wave function) that the spin of an electron in the first part of the equipment is either up or down. And if we imagine that the ball can be analyzed through the interconnections up to this dimension (1 cm) by the quantum mechanics, then before we make an observation we still have to give an amplitude that the ball is up and an amplitude that the ball is down.

This is the standard von Neumann measurement and registration device,[7] which connects a microscopic variable unitarily with macroscopic ones. In contrast to Belinfante, Feynman is here evidently using "amplitudes" (wave functions) rather than operators as the dynamical objects of the theory. This description of a measurement is closer to that of the "Princeton school" (von Neumann, Wigner) than to Copenhagen (Heisenberg or Bohr, who would both have used observables and classical variables at some point). Note that the famous formal "equivalence" between the Schrödinger and the Heisenberg picture is well defined and unproblematic only for closed systems, where the Hamiltonian that has to be used for the transformation between "pictures" would not affect the environment.

> "Now, since the ball is big enough to produce a *real* gravitational field (we know there's a field there, since Coulomb measured it with a 1 cm ball), we could use that gravitational field to move another ball, and amplify that, and use the connections to the second ball as the measuring equipment.

Shifting the Heisenberg cut. It is interesting that according to recent historical studies,[8,9] Heisenberg and Bohr seem to have differed about its precise meaning. While Heisenberg insisted on the free variability of its position anywhere between object and observer, Bohr placed it at the end of the quantum measurement proper (the first one in Feynman's chain of measurements or interactions) that in his opinion would create objective classical values. In the following, Feynman sticks to further tradition in neglecting the unavoidable environment in his description.

> We would then use that gravitational field to move another ball, and amplify that, and use the connections to the second ball as the measuring equipment. We would then have to analyze through the channel provided by the gravitational field itself via the quantum mechanical amplitudes.
>
> "Therefore, there must be an amplitude for the gravitational field,

This formulation is remarkable, since (1) his "must *be*" already indicates some ontic interpretation of the wave function, and (2) it refers to a wave functional (a Schrödinger picture for fields – in distinction to time-dependent field operators carrying the dynamics). This will be further illustrated below. Ten years later, the concept of wave functionals for gravitational fields led to the Wheeler-DeWitt equation[10,11] – in spite of its technical and interpretational problems the only conventional ("canonical") quantization of gravity as an empirically founded "effective" quantum theory that does not add any speculative novel elements.

> *provided* that the amplification necessary to reach a mass which can produce a gravitational field big enough to serve as a link in the chain, does not destroy the possibility of keeping quantum mechanics all the way. There is a *bare* possibility (which I shouldn't mention!) that quantum mechanics fails and becomes classical again when the amplification gets far enough, because of some minimum amplification which you can get across such a chain.

So Feynman considers a modification of the Schrödinger equation or any other limitation of quantum mechanics as bare possibilities that should not even be mentioned! Note that this problem here logically precedes the question whether *gravity* has to be quantized or not.

> But aside from that possibility, if you believe in quantum mechanics up to any level then you have to believe in gravitational quantization in order to describe this experiment.
>
> "You will note that I use gravity as part of the link in a system on which I have not yet made an observation. The only way to avoid quantization of gravity can *in principle* no longer play a role beyond a certain point in the chain, and you are not allowed to use quantum mechanics on such a large scale. But I would say that this is the only 'out' if you don't want to quantize gravity."

In this part of the discussion, Feynman seems to consider the observer as the ultimate link in the chain that must lead to a unique measurement result. This is again tradition. It corresponds to Heisenberg's early idealistic concepts as well as to von Neumann's and Wigner's "orthodox" interpretation – but not to Bohr's one, who would presume objective classical concepts to describe the pointer states ("indicators") of a measurement device. The question is then only, *where* unitarity would break down.

> **Bondi:** "What is the difference between this and people playing dice, so that the ball goes one way or the other according to whether they throw a six or not?
>
> **Feynman:** "A very *great* difference. Because I don't really have to measure whether the particle is here or there. I can do something else: I can put an inverse Stern-Gerlach experiment on and bring the beams back together again. And if I do it with great precision, then I arrive at a situation which is not derivable simply from the information that there is a 50 percent probability of being here and a 50 percent probability of being there. In other words, the situation at this stage is *not* 50-50 that the die is up or down, but there is an *amplitude* that it is up and an *amplitude* that it is down – a *complex* amplitude – and as long as it is still possible to put those amplitudes together for interference you have to keep quantum mechanics in the picture.

This is the standard argument against an epistemic interpretation of the wave function (so he says "*there is* an amplitude"). It also excludes an interpretation of the Feynman path integral as representing an ensemble of "potential" paths. The reduction of the wave function can thus *not* be regarded as a mere increase of information. In fact, Feynman's remarks in these conference proceedings seem to have later caused Roger Penrose to suggest a gravity-induced collapse as a modification of the Schrödinger equation.[12] It is remarkable that Feynman has here to repeat this well-known argument, but the insufficient, merely statistical interpretation of the wave function is still very popular today, since it is convenient for describing the situation *after* an irreversible measurement. It is used in most textbooks, and usually expected as an answer from physics students in their examination. Feynman's last half-sentence seems to allow for decoherence as a solution of the problem, since a macroscopic gravitational field (precisely as its macroscopic source) is permanently being "measured" by environmental particles[13,14] – but because of his further arguments I doubt that he would have accepted this explanation as a complete one.

> "It may turn out, since we've never done an experiment at this level, that it's not possible – that by the time you amplify the thing to a level where the gravitational field can have an influence, it's already so big that you can't reverse it – that there is something the matter with our quantum mechanics when we have too much *action* in the system, or too much mass – or something.

He is again talking about a real collapse process as a modification of the Schrödinger equation – not of environmental decoherence.[15]

> But that is the only way I can see which would keep you from the necessity of quantizing the gravitational field. It's a way that I don't want to propose. But if you're arguing legally as to how the situation stands …"
>
> **Witten:** "What prevents this from becoming a practical experiment?"
>
> **Feynman:** "Well, it's a question of what goes on at the level where the ball flips one way or the other.

Or when Schrödinger's cat dies!

> In the amplifying apparatus there's already an uncertainty – loss of electrons in the amplifier, noise, etc, – so that by this stage the information is completely determined. Then it's a *die* argument.
>
> "You might argue this way: Somewhere in your apparatus this idea of amplitudes has been lost. You don't need it any more, so you *drop* it. The wave packet would be reduced (or something). Even though you don't know *where* it's reduced, it's reduced. And then you can't do an experiment which distinguishes *interfering* alternatives from just plain odds (like with dice).

Sounds much like decoherence. But wait – he has not yet made clear what exactly he means!

> "There is certainly nothing to prevent this experiment from being carried out at the level at which I make the thing go 'clink-clank', because we do it every day: We sit there and we wait for the count in the chamber – and then we publish, in the Physical Review, the information that we've obtained *one pi meson*. And then it's printed (bang!) on the printing presses – stacked and sent down to some back room – and *it moves the gravitational field!*
> 
> "There's no question that if you have allowed that much amplification you have reduced the wave packet. On the other hand it may be that we can think of an experiment – it may be worth while, as a matter of fact, to try to design an experiment where you can invert such an enormous amplification."
> 
> **Bergmann:** "In other words, if it is established that nobody reads the Physical Review, then there is a definite 50 % uncertainty …"
> 
> **Feynman:** "Well, some of the copies get lost. And if some copies get lost, we have to deal with probabilities again."

Here the "other copies" are indeed used as an uncontrollable (though macroscopic) environment, even though entanglement (which is responsible for quantum decoherence) is not mentioned. So he overlooks the unavoidable microscopic environment (the other copies could simply be replaced by thermal photons or molecules, for example). Completely inverting the amplification process in practice would require the coherent return of all chaotically scattered particles.

> **Rosenfeld:** "I do not see that you can conclude from your argument that you must quantize the gravitational field. Because in this example at any rate, the quantum distinction here has been produced by other forces than gravitational forces."
> 
> **Feynman:** "Well, suppose I could get the whole thing to work so that there would be some kind of interference pattern. In order to describe it I would want to talk about the interaction between one ball and the other. I could talk about this as a direct interaction like $\psi^2/r_{ij}$. (This is related to the discussion of whether electrostatics is quantized or not.) However, if you permit me to describe gravity as a field then I must in the analysis introduce the idea that the field has *this* value with a certain *amplitude*, or *that* value with a certain *amplitude*.

Feynman is again using the superposition principle as the essential aspect of quantization – but neither the uncertainty principle nor any nonclassical algebra of formal "observables" which would represent uncertain *classical* variables.

> This is a typical quantum representation of a field. It can't be represented by a classical quantity. You can't say what the field *is*. You can only say that it has a certain amplitude to be this and a certain amplitude to be that, and the amplitudes may even interfere again … possibly. That is, if interference is still possible at such a level."

Does this not mean (for Feynman, too) that quantum amplitudes – in contrast to classical fields – represent *real* properties in any reasonable sense (not just probabilities for something else)?

> **Rosenfeld:** "But what interferes has nothing to do with gravitation."

Rosenfeld is right. This is a general discussion of classicality – not just of quantum gravity. This confusion may also later have misled Penrose to relate his collapse proposal to gravity. Bohr would neither have been happy with amplitudes for balls nor for gravitational fields.

> **Feynman:** "That's true … when you finish the whole experiment and analyze the results. *But*, if we analyze the experiment in time by the propagation of an amplitude – saying there is a certain amplitude to be here, and then a certain amplitude that the waves propagate through there, and so on – when we come across this link – if you'll permit me to represent it by a gravitational field – I must, at this stage in time, be able to say that the situation is represented now *not* by a particle here, *not* by a result over there, but by a certain amplitude for the field to be this way and a certain amplitude to be that way. And if I have an amplitude for a field, that's what I would define as a quantized field."

In order to understand Feynman's cumbersome arguments to answer Rosenfeld, one has to remember that the need to quantize even the electromagnetic field was still questioned long after this debate – until lasers and cavity electrodynamics became available. In contrast to quantum fields, classical forces-at-a-distance would not lead to decoherence before causing any effects on quantum matter, for example, since forces would not be "traced out".

> **Bondi:** "There *is* a little difficulty here (getting onto one of my old hobby horses again!) if I rightly understand this, which I'm not sure that I do: The *linkage* must not contain any irreversible elements. Now, if my gravitational link radiates, I've had it!"
> **Feynman:** "Yes, you have had it! Right. So, as you do the experiment you look for such a possibility by noting a decrease of energy of the system. You only take those cases in which the link doesn't radiate. The same problem is involved in an electrostatic link, and is not a relevant difficulty."

Real (contrasted to virtual) decoherence is now known to be the most efficient irreversible process in Nature.[16] However, Bondi and Feynman here argue in terms of classical irreversi-bility (radiation and energy loss) – not in terms of an irreversible spread of phase relations and entanglement (dislocalization of superpositions). The paper by Feynman and Vernon, which *could* have described decoherence, would appear in 1963,[17] but its authors applied it only to microscopic degrees of freedom – not to explain classicality – and they did not appropriately distinguish decoherence from dissipation. (As Wojciech Zurek once told me, Feynman became very interested in the concept of decoherence shortly before his death.)

The discussion of whether radiation effects can be avoided now continues:

> **Bondi:** "Oh yes, because in the electrostatic case I can put a conducting sphere around it …"
>
> **Feynman:** "It doesn't make any difference if it radiates. If every once in a while the particle which is involved is deflected *irreversibly* in some way, you just remove those cases from your experiment. The occurrence could be observed by some method outside."

This "removal" might already require a collapse, however.

> **Bergmann:** "Presumably the cross section for gravitational radiation is extremely …"
>
> **Feynman:** "And *furthermore*, we can estimate what the odds are that it will not happen."
>
> **Bondi:** "I am just trying to be difficult."
>
> **Gold:** "Well, it could still be that some irreversible process is necessarily introduced by going to a thing as big as that."
>
> **Feynman:** "Precisely what I said was the only way out."
>
> **Gold:** "But that need not mean that there is some profound thing wrong with your quantum theory. It can mean merely that when you go into the details of how to make an op. …"

Gold here applies an argument that has very often been used in attempts to solve the measurement problem: introduce sufficient complications which appear similar to a complex classical amplification process. However, the linearity argument which leads to Schrödinger's cat superposition is impeccable within quantum theory (as has been explained and emphasized by von Neumann, Wigner, and others). So Feynman counters:

> **Feynman:** "There would be a *new principle!* It would be *fundamental!* The principle would be – *roughly: Any piece of equipment able to amplify by such and such a factor* ($10^{-5}$ grams or whatever it is) necessarily *must be of such a nature that it is irreversible*.

Here he evidently refers again to a fundamentally irreversible collapse of the wave function as a modification of the Schrödinger equation.

> "It might be true! But at least it would be fundamental because it would be a new principle. There are two possibilities. Either this principle – this missing principle – is right, *or* you can amplify to any level and still maintain interference, in which case it's absolutely imperative that the gravitational field is quantized … *I believe! Or* there's another possibility which I haven't thought of."

Quantum gravity, which was the subject of the discussion, appears here only as a secondary consequence of the assumed absence of a collapse, while the major one is that "interference" (superpositions) must always be maintained according to quantum theory. There is hence *no ensemble* of possible states that would represent incomplete knowledge as for the die. Because of Feynman's last arguments it is remarkable that nobody here mentioned Everett's ideas, since some of his early drafts must have been known to some participants at this time, and his thesis would be accepted by Reviews of Modern Physics just a few weeks after the conference.[18] Feynman himself seems to have known it already, as he referred to the "universal wave function" when it was mentioned by John Wheeler in Session 9 – see below.

> **Buckingham:** "The second possibility lands you back in the same difficulty again. If you *could* amplify to any factor, you could reduce to a negligible proportion an additional signal to take an observation on, say, those balls."
>
> **Feynman:** "No!"
>
> **Buckingham:** Because you only need one light quantum."
>
> **Feynman:** "No!"
>
> **Buckingham:** "If you could amplify up to any factor this becomes negligible."

Decoherence arguments would immediately have proven this statement wrong. For Feynman it does not appear that easy.

> **Feynman:** "It depends! … You see [pointing to a blank space on the blackboard] this statement that I have written here is not written very precisely – as a matter of fact if you look at it you probably can't even see the words. I haven't thought out how to say it properly. It isn't simply a matter of amplifying to any factor. It's too crude – I'm trying to feel my way.

This simple analogy is remarkable: a property to be measured is either initially absent and must hence be "created" in the measurement, or the measurement is reversible in principle. A reversal of decoherence would require a recombination of all "branches" of the universal wave function.

> We *know* that in any piece of apparatus that has ever been built it would be a phenomenally difficult thing to arrange the experiment so as to be reversible. But is it *impossible*? There is nothing in quantum mechanics which says that you can't get interference with a mass of $10^{-5}$ gram – or *one* gram."
>
> **Buckingham:** Oh, yes. What I'm saying, though, is that the laws have to be such that the effect of one light quantum is sufficient to determine which side the ball is on, and would be enough to disturb the whole experiment."

Here is the typical confusion between decoherence by uncontrollable entanglement and a *distortion* of the considered system that could indeed be neglected on the level of a single photon. But Feynman does not see that point either:

> **Feynman:** "Certainly! That's always true. That's just as true no matter what the mass is."
>
> **Anderson:** "Suppose a neutral elementary particle really has a gravitational field associated with it which you could actually use in the causal link. The thing that bothers you is that you may be getting something that is too small to produce a gravitational field."
>
> **Feynman:** "It's a question of design. I made an assumption in this analysis that if I make the mass too small the fields are so weak I can't get the experiment to operate. That might be wrong too. It may be that if you analyze it close enough, you'll see that I can make it go through a gravitational link without all that amplification – in which case there is no question. At the moment all I can say is that we'd better quantize the gravitational field, or else find a new principle."

A similar problem has survived in decoherence theory: at what mass difference is a superposition of two different masses decohered by their own gravitational field in analogy to charge superselection rules?[19] Evidently, there *are* time-dependent quantum states, which must be superpositions of slightly different energies – in conflict with an *exact* analogy. So this seems to be a quantitative question that has not yet been sufficiently analyzed.

> **Salecker:** "If you assume that gravitation arises as a sort of statistical phenomenon over a large number of elementary particles, then you also cannot perform this experiment."
>
> **Feynman:** "Yes, it depends what the origin is. One should think about designing an experiment which uses a gravitational link and at the same time shows quantum interference – what dimensions are involved, etc. Or if you suppose that every experiment of this kind is impossible to do, you must try to *state* what the general principle is, by trying a few examples. But you have to state it right, and that will take some thinking.

Since the time of the conference, many collapse mechanisms and corresponding experiments have been designed – some of them based on gravity. No deviation from unitarity has ever been confirmed, while quantum interference has been demonstrated for pretty large systems.

**3. Some remarks from Section 9**

Session 8 continued with some further comments on the quantization of gravity. Toward the end of the conference (in the final Session 9), **John Wheeler** mentioned that

> there exists another proposal that there is one "universal wave function". This function has already been discussed by Everett, and it might be easier to look for this "universal wave function" than to look for all the propagators.

We now understand that, for most systems, separate propagators would not even exist because of their unavoidable entanglement (which locally leads to decoherence).

> **Feynman** said that the concept of a "universal wave function" has serious conceptual difficulties. This is so since this function must contain amplitudes for all possible worlds depending on all quantum-mechanical possibilities in the past and thus one is forced to believe in the equal reality [sic!] of an infinity of possible worlds.

Well said – although we may restrict ourselves *in practice* to an individual autonomous branch of this wave function ("our world"). The total number of such branches must indeed be extremely large, but it need not be infinite, since coherence lengths for continuous variables never vanish exactly. Reality *is* conceptually difficult and complex, and we should not be surprised that it seems to go far beyond what *we* will ever be able to observe. But Feynman is not ready to draw this ultimate conclusion from the superposition principle that he always defended during this discussion. Why should a superposition not be maintained even when it includes an observer? Why "is" there not an amplitude for me (and you) observing this and an amplitude for me (and you) observing that in a quantum measurement – just as it would be required by the Schrödinger equation for a gravitational field?

> Quantum amplitudes represent more than just probabilities. Bondi's question "What is the difference to people playing dice" probably represents the most frequent misunderstanding of quantum theory. Feynman answered it by using an inverse Stern-Gerlach device. This experiment would not be possible any more if the two wave packets into which the wave function has split had been irreversibly decohered from one another, as it must always happen when they are measured by means of a macroscopic device. Unless the unitary dynamics were changed, this would necessarily lead to different autonomous "worlds". But even if there was something like a stochastic collapse occurring *after* decoherence, Feynman's argument could still be applied backwards in time by using the original Stern-Gerlach device. For a die in a deterministic world, one would assume the existence of different unknown causes for its different final states, while the past of the Stern-Gerlach experiment is quantum mechanically described by *one and the same* individually meaningful superposition of both outcomes (the same superposition that would indeed be recovered by means of the inverse device in the future when used before an irreversible measurement had been performed). So the quantum situation can *not* be explained in terms of lacking information about any causes (an initial ensemble), while the modernistic talk about some fundamental "quantum information" does not explain anything, remains purely verbal, and only adds to the confusion.

Feynman then gave a resumé of the conference, adding some "critical comments", from which I would like to quote a remark addressed to mathematical physicists (page 150):

**Feynman:** "Don't be so rigorous or you will not succeed."

(He explains in detail how he means it.) It is indeed a big question what mathematically rigorous theories can tell us about reality, since their axioms can never be *exactly* empirically founded (as emphasized already by Henry Poincaré[20]). This question is particularly pressing if the formal theory does not even contain the most general axiom of quantum theory: the superposition principle. So towards the end of his resumé (page 154), he said

**Feynman**: "Even if one believes in exhausting the classical problems first and also believes in unification, there is some question as to whether the unified theories are the correct starting point. At the end of the list of classical problems there is the real problem of the feasibility of separating the strictly classical questions from the quantum questions."

Doesn't this warning perfectly apply to the present search for a unified theory? The important lesson from decoherence theory was that the superposition principle holds even where it did not seem to hold, and that classical concepts emerge from a universal *quantum* theory: their superpositions cannot be observed locally even if they persist in the global wave function. Nonetheless, many modern field theorists and cosmologists seem to regard quantization as of secondary or merely technical importance (just providing certain "quantum corrections") for their endeavors, which are essentially based on classical terms – such as classical fields – (see also Ch. 6 of Ref. 16). It is then not surprising that the measurement problem never comes up to them. How can anybody even argue about unified quantum field theories or cosmology (which must both *include* a description of observers) without first defining his interpretation, that is, without clarifying whether he/she is using Everett's interpretation or some kind of collapse mechanism (or something even more speculative than a collapse)?

**Acknowledgment:** I wish to thank Claus Kiefer for drawing my attention to the Chapel Hill report, and in particular to Feynman's remarks on the meaning of quantization. I am also grateful to Charles Misner for correcting two of my comments concerning the relation of the discussion to Everett's work in an early version of the manuscript, and to Wolf Beiglboeck for suggesting some additional comments.